\documentclass{amsart}%
\usepackage{amssymb,graphicx,color}

\def\sz{\char'31}%
\def\etc{{\it etc.\/}}%
\def\ie{{\it i.e.\/}}%
\def\eg{{\it e.g.\/}}%
\def\Eg{{\it E.g.\/}}%
\def\vs{{\it vs.\/}}%
\def\BbbR{\mathbb R}%
\def\hatp{\hat p}%
\def\bfn{\mathbf n}%
\def\bfp{\mathbf p}%
\def\bfE{\mathbf E}%
\def\bfhatp{\mathbf {\hatp}}%
\def\bl{\bigl}%
\def\br{\bigr}%
\def\Bl{\Bigl}%
\def\Br{\Bigr}%
\def\Pr{{\mathbf{Pr}}}%
\def\eps{\varepsilon}%

\definecolor{darkgreen}{rgb}{0,.6,0}
\definecolor{darkagenta}{rgb}{.5,0,.5}
\definecolor{darkred}{rgb}{1,0,0}
\definecolor{darkblue}{rgb}{0,0,.4}
\definecolor{black}{rgb}{0,0,0}

\newif\ifShowComments
\ShowCommentstrue
\def\strutdepth{\dp\strutbox}
\def\druk#1{\strut\vadjust{\kern-\strutdepth
      {\vtop to \strutdepth{%
	      \baselineskip\strutdepth\vss
		      \llap{\hbox{#1}\quad}\null}}}}
\def\asksign{{\bf?!!}}

\def\commO#1{\ifShowComments\hfill\\\asksign\hfill%
\\\druk{{\Large\bf?!! O.H.}}\kern-.45em{\bf\ #1 }\par\noindent%
\asksign\par\noindent\kern-0em\fi\\}

\def\writefig#1 #2 #3 {\rlap{\kern #1 ex \raise #2 truecm 
\hbox{\small#3}}}
\def\figtext#1{\smash{\hbox{#1}}\vspace{-5mm}}
\begin{document}

\markboth{A.Bovier, J.\v Cern\'y, O.Hryniv}
{Opinion Game}

%

\title[Opinion Game]
{The Opinion Game: 
 Stock price evolution from microscopic market modelling}
\author{Anton Bovier}
\address{Weierstra\sz{}--Institut
f\"ur Angewandte Analysis und Stochastik\\
Mohrenstrasse 39, 10117 Berlin, Germany \and\
Institut f\"ur Mathematik, Technische Universit\"at Berlin\\
Stra\sz e des 17. Juni, 136, 10623 Berlin, Germany}
\email{bovier@wias-berlin.de}

\author{Ji\v r\'\i\ \v Cern\'y}
\address{Weierstra\sz{}--Institut
f\"ur Angewandte Analysis und Stochastik\\
Mohrenstrasse 39, 10117 Berlin, Germany}
\email{cerny@wias-berlin.de}

\author{Ostap Hryniv}
\address{Statistical Laboratory, Centre for Mathematical Sciences\\
University of Cambridge, Wilberforce road, Cambridge CB3~0WB, UK}
\email{o.hryniv@statslab.cam.ac.uk}

\begin{abstract}
We propose  a class of Markovian agent based models for the 
time evolution of a share price in an interactive market. The models
rely on a microscopic description of a market of buyers and sellers
who change their opinion about the stock value in a stochastic way.
The actual price is determined in  realistic way by matching
(clearing) offers until no further transactions can be performed.  
Some
analytic results for a non-interacting model
are presented. We also propose  
basic interaction mechanisms and show in  simulations that these
already reproduce 
certain
particular features of prices in real stock markets.
\end{abstract}


\maketitle

\section{Introduction}

The financial markets constitute an intriguing and complex system that
has not failed to attract mathematicians and scientists from other
fields 
for a long time. Only rather recently, however,
has mathematical finance, and more specifically the theory of
derivatives on the stock markets become a major field of mathematics
and one of the major sources of inspiration for probability theory in
general and stochastic analysis in particular. 
The reason for this development is simple: It is based on the apparent 
success of the so-called Black-Scholes formula for the fair price of an option
as a tool for the actual trader on the market. Indeed, the very existence of 
this mathematical theory appears to be largely responsible for the recent 
growth and diversification of the derivative market itself, which in its 
present form would have been impossible without an underlying mathematical
theory.  On the other side, the great success of this same theory in the 
mathematical sciences is due, beyond the obvious advantages it provides 
for the careers of students trained in this field, largely to the fact that
there is a very clear mathematical setting for this theory with clearly 
spelled out axioms and assumptions which allows  for the 
mathematician to bring his traditional weapons to bear in a familiar terrain.

One of the crucial issues in the financial mathematics is the
modeling of prices of commodities (stocks, currencies \etc) with help
of stochastic processes. The main approaches that have been used in
this context are the following: 
\begin{itemize}
\item {\it generalized Black-Scholes (BS) theory.} Originally the BS
theory \cite{BS} (for good textbook exposition see, \eg, \cite{KS,FS})
emerged as a theoretical foundation of pricing of derivatives
(options) of underlying financial instruments (stocks, 
currencies \etc). Initially the price of the underlying was taken as
geometric Brownian motion. Later this theory was generalized and put
on the axiomatic background based on the assumption of the so-called
non-arbitrage condition. This led to the conclusion that prices are
described by semi-martingale measures, and thus are essentially given
by solutions of certain stochastic differential equations.
This
framework has been the main driving force of the rapid growth of
financial mathematics in the last decades. Note that the main purpose
of this theory is not to derive specific models for the underlying
but rather to deduce consequences from generally accepted
principles. In this respect this approach can be compared to
classical thermodynamics. 

\item {\it statistical approach.} 
Empirical studies of share price data try to model the data by certain 
stochastic process. Popular classes of models are ARCH, GARCH, ARMA \etc\
There seem to be no totally conclusive results, 
but certain interesting phenomena have been observed, such as
universal exponents in certain correlation data (see \cite{B-N}). Quite
frequently, analogies to phenomena like turbulence are drawn.  
This approach does not  usually 
intend to derive the model from any underlying
economic theory.

\item {\it agent based models.} ``As prices are generated by the
  demand of agents who are active on the financial market for the
  given asset, a [\dots] model [\dots] should be explained in terms
  of the interaction of these agents.'' Based on this observation
  stated by F\"ollmer in \cite{F}, a large number of agent based
  models for price evolution have been developed. F\"ollmer in
  \cite{F} suggested a model of diffusion in random environment by
  viewing the price process as a sequence of temporary equilibria in
  a market with agents. 
  Among the most popular microscopic models for financial markets are
  so-called agent based models that are more or less sophisticated
  versions of the ``minority game'' (MG); for reviews, see Jeffries
  and Johnson \cite{JJ}, Bouchaud and Giardina \cite{GB}, and
  references therein. In all these models there exists a collection
  of traders, each endowed with the possibility to make a decision
  (typically of the type ``buy'', ``sell'', or ``hold'') concerning a
  given investment. To reach such a decision, the trader disposes of
  a certain number of ``strategies'', a strategy typically being a
  function of the price history into the set of decisions. The game
  consists of the traders choosing their strategies in a way as to
  reach certain objectives (in the minority game to be ``in the
  minority'', in reality to make a maximal profit from the
  transactions undertaken). At each time step, the asset price is
  updated according to an empirical rule as a function of  the number
  of ``buyers'' and ``sellers''. There are currently a large number
  of versions of such models around, including models with additional
  stochastic components. These models exhibit a rather rich dynamical
  structure. However, they are rather heavy handed both analytically
  (where little or nothing is known on a mathematical level) and
  numerically. Moreover, the large number of assumptions and
  parameters entering the models makes their predictive power
  somewhat limited. Purely deterministic models of this type have
  been criticized before (Bouchaud and Giardina \cite{GB}) and
  stochastic models have been proposed that should allow to take into
  account irrational behaviour of agents. 
\end{itemize}

In this paper we want to propose agent based market models that are 
{\it much simpler} and that, at the same time, are build reasonably close to 
what is actually happening in financial markets. 
The main distinction between 
our model and the MG type models is that it focuses on collective effects of 
a market while not attempting to model the actual reasoning process of an
individual trader. 
The basic paradigm of our modeling approach is the 
notion of a price. Prices of a share of stock or other 
commodity arise from trading. There are various 
developed theories in economics concerning prices based on some equilibrium 
assumptions, but, fortunately, in the stock market in particular,
the price of a share is obtained by a well defined procedure which is easily 
implementable in an algorithm provided a sufficient amount of
information about the state of the market at any given time is available.
The basic principle of our approach is the modeling of a set of
interacting agents in a way that allows to extract the price from the
current trading state in a rigorous way.

The remainder of this article is organized as follows. In
Section~\ref{two} we explain the basic principles of our modeling 
approach in detail and illustrate them in some special cases.
In Section~\ref{three} we analyze
 some of the phenomena that can
already be observed in this simplest setup and relate them to certain
features of real world data. In Section~\ref{four} we introduce
a number
of additional features that should be implemented to obtain more
realistic models, and discuss
 some of the additional effects they 
produce.
In Section~\ref{five} we present our conclusions.

\section{Our approach\label{two}}
The basic procedure of price determination in real stock markets is
done by trading. That is to say, participating traders propose prices
at which they are willing to sell respectively buy a certain number of
shares of a given stock. The list of these offers at a stock exchange
is called the orderbook. On the basis of this orderbook the market
maker is matching buy and sell offers according to certain rules until
no further transactions are possible. That is to say, after the
matching procedure (clearing) the highest price proposed for buying a
share is smaller than the lowest price at which a share is
offered. These two values are then quoted as the bid price and the ask
price. The dynamics of the real stock market thus has two components:

\begin{itemize}
\item the changing of buy and sell offers made by traders and
\item the matching of these offers by the market maker which fixes the
quoted price at a given time.
\end{itemize}
Our purpose is to develop a class of models which reflects this
mechanism of pricing and allows for diverse modeling on different
levels of complexity of the behaviour of the agents, while maintaining
the pricing mechanism by the market maker. To do so, a minimal
requirement for the description of the state space of the trading
agents is that it must allow us to recover the state of the orderbook
at any given moment in time. 

Our idea is thus to consider the time evolution of a virtual orderbook
or a ``trading state''
containing the opinions  of each participating agent about the
``value''%
\footnote{We will distinguish the notion of the value from that of the
price. The {\it value} is what agents have an opinion about, while
the {\it price} is determined by the market.
The
opinion on the value can be driven by fundamental considerations
(\eg\ earning or  dividend expectations, typically coming from
outside information), or speculative considerations
(\eg\ predictions based on partial knowledge on the current state of
the opinions of other traders), or both.}  
of the stock. The evolution is driven by the change in opinion of the
agents and the action of the market maker.

A minimal model in which this idea can be implemented can be described
as follows. We consider trading in one particular stock.
Assume that there are $N$ ``traders'' and $M<N$ shares of the
stock. We make the simplifying assumption that each trader can own at
most one share. The state of {\it each } trader $i$ is given by its opinion
$p_i\in\BbbR$ about the logarithm of the value of the stock, and by
the number of shares he owns $n_i\in \{0,1\}$. This is
to say,
the trader $i$ would be willing to sell his share at the
price $e^{p_i}$, if he owns one ($n_i=1$), respectively buy a share at
this price, if he does not own one ($n_i=0$). We say that a trading
state is \emph{stable}, if the 
$M$ traders having the $M$ highest opinions $p_i$ all own a
share. 
This means in particular that in a stable state one can infer
the set of owners of shares from the knowledge of the  state of opinions
$\bfp=(p_1,\dots,p_N)$. 
Thus a stable trading state is completely
determined by the set of $N$ values $p_i$, and we will in the sequel
identify stable trading states with the vector $\bfp$. As we will
normally only work with stable trading state, we suppress the qualifier
stable when no confusion can arise. 

Given a stable trading state $\bfp$ 
we denote by
$\bfhatp=(\hatp_1,\dots,\hatp_N)$ its order statistics, that is $\hat
p_i=p_{\pi_i}$ for a permutation $\pi\equiv\pi(\bfp)$ of the set of
$N$ elements such that $\hatp_1\le\dots\leq\hatp_N$. Then the number
of shares owned by traders,
$\bfn(\bfp)=\bl(n_1(\bfp),\dots,n_N(\bfp)\br)$ satisfies
\[
n_i(\bfp)=
\left\{
\begin{array}{ll}
0, & \hbox{ if $\pi_i(\bfp)\le N-M$,}\\
1, & \hbox{ if $\pi_i(\bfp)\ge N-M+1$.}
\end{array}
\right.
\]
To a trading state $\bfp$ we associate the {\it ask price\/} 
\[
p^a\equiv p^a(\bfp)= \hatp_{N-M+1}\,,
\]
and the {\it bid price\/}
\[
p^b\equiv p^b(\bfp)= \hatp_{N-M}\,.
\]
Obviously, $p^a(\bfp)$ is the lowest price asked by traders owning a
share and $p^b(\bfp)$ is the highest price offered by traders wanting
to buy a share. For convenience we will refer to $\frac 12(p^a+p^b)$
as the {\it current price} in the sequel.

Any dynamics $\bfp(t)$ defined on the trading state induces the
dynamics of $\bfhatp$ and in particular of the pair
$\bl(p^a,p^b\br)$.

Our next simplifying assumption is that the above trading state $\bfp$
evolves in time as a (usually time-inhomogeneous) Markov chain%
\footnote{We will discuss the Markovian assumption later.} 
$\bfp(t)$ with state space $\BbbR^N$. We will further assume that time
is discrete 
(this is inessential but more convenient for computer simulations) 
and that the updating proceeds asynchronously, \ie\ typically at a given
instant only a single opinion changes. This dynamics can be considered
as an interacting particle system, however,  some special features
should be incorporated that reflect the peculiarities of a market.

The first and most obvious one is that the ask and bid prices
$p^a(t),p^b(t)$  are likely to be important quantities which will
influence the updating probabilities.

Moreover, it is reasonable to distinguish between transitions that
leave $\bfn$ unchanged, and those for which $\bfn\bl(\bfp(t+1)\br)\neq
\bfn\bl(\bfp(t)\br)$.
In the latter case we say that a {\it transaction\/} has occurred.

Before we discuss some more specific implementations of this general
setup a few remarks  concerning some features that may appear
offending are in place. 

The first is doubtlessly the assumption that the process is
Markovian. This appear unnatural because the most commonly available
information on a stock is the history of its price, the ``chart'', and
most serious traders will take this information at least partly into
account when evaluating a stock, with some making it the main basis
for any decision.%
\footnote{A frequently heard remark being that all information on a stock 
is in its price.}
Certainly one could retain such information and formulate a
non-Markovian model, as \eg\ the model of Bouchaud and Giardina
\cite{GB}. However, if one starts to think about this, one soon finds
that it is very difficult to formulate reasonable transition rules on
the basis of the price history. On a more fundamental level, one will
also come to the conclusion that the analysis of the history of the
share price is in fact performed in order to obtain information of the
{\it current opinion of the traders concerning the value of the stock}
with the hope of inferring information on the future development of
the price. For instance, if one knew that there are many people willing
to sell shares at a price not much higher than $p^a$, one knows that
it will be difficult for the price to break through this level (this
is know as a ``resistance'' by chart-analysts and usually inferred
from past failures to break through such a level). Therefore, instead
of devising rules based on past price history, we may simply assume
that the market participants have some access to the prevailing
current opinions, obtained through various sources
(chart analysis, rumours, newspaper articles, \etc) and take this into
account when changing their own estimates. 

The second irritating point is that money does not appear in our model except 
in the form of opinions about values. In particular, we do not keep track of 
the cashflow of a given investor (that is to say we do not care whether a 
given investor wins or looses money).  There are various reasons to justify 
this. First we consider that the market participants do not invest a 
substantial fraction of their assets in this one stock, so that shortage of 
cash will not prevent anyone to buy if she deems opportune to do so 
(in the worst case money can be obtained through credits). Then, money is not 
conserved, but the total value of the stock can inflate as long as there is 
enough confidence. Also,  we do not keep track of the objective success 
of a trader,  because we do not know how this will eventually influence her 
decisions. While a given trader may follow her personal strategy with the
hope of making profits, we cannot be sure that these strategies will succeed.
What is important and what is built into our model, however, is the fact that
any trader%
\footnote{
We could enlarge the model to incorporate a small fraction
of traders which do not act according to common sense, or against their own
convictions (\eg\ traders that have bought their stock on credit and are
executed by their creditors on falling prices. It may be interesting
to consider the effect of that in the context of market crashes).} 
will have the subjective impression to make a profit at any
transaction.%
\footnote{Which also implies that this subjective opinion must be wrong at
least for one of the traders involved. But this seems to reflect
reality.} 
Thus we feel that opinions about values are the correct variables to
describe such a market rather than the actual flow of capital, at
least at the level of a simple model. 

The above setting  suggests a rather general and flexible class of
models of a stock market. Its main feature is that it describes the
time evolution of a share price as the result of an interacting random
process that reflects the change of the opinions of individual traders
concerning the value of the stock. Even when this last process is
modelled as a Markov process, the resulting price process
$\bl(p^a(t),p^b(t)\br)$ will in general not be a Markov process.

\section{Examples\label{three}}

\subsection{
Ideal gas approximation}

Obviously the simplest model for the dynamics of the trading state 
$\bfp(t)$ is to choose it as a collection of  independent identically 
distributed one-dimensional Markov processes (``random walks'') 
$p_i(t)$. This corresponds to the ideal gas approximation in 
statistical mechanics. In this case, the price process is simply 
obtained from the order statistics of independent processes and 
asymptotic results for $M$ and/or $N$ large
(recall that $M$ denotes the 
    number of traded shares and $N$ the total number of traders) can 
be obtained rather easily. While this model is somewhat
simplistic, some rather interesting phenomena can already be modelled 
in this context, as we will explain now.

We may be interested in a situation where some macro-economic model
may predict several stable (respectively metastable) values of the stock price,
realized as the minimum of some utility function $V$.

In such a situation it seems not unreasonable to model the 
process of a single trader as a one-dimensional 
diffusion process with drift obtained from a
potential function $V$, \ie\
we can take $p_i(t)$ to be a solution of the stochastic differential 
equation 
\[
dp_i(t) = -V'\bl(p_i(t)\br) dt +\sqrt \eps\, dW_i(t)
\]
with  $W_i(t)$ i.i.d.\ standard Brownian motions, and $\eps>0$ a parameter
measuring the diffusivity. Alternatively,
 we can take discrete approximations
of this process, as will always be done in numerical simulations.

Let us consider the situation when there are two (meta) stable values
of the price,  $q_1$ and $q_2$, \ie\ the situation where the potential 
$V$ has two minima (wells) at $q_1$ and $q_2$. If the potential is strong,
resp.\ $\eps  $ is small, an individual trader would typically spent
long periods of time near one of the favoured values $q_1$ or $q_2$. 

Let $w_1$ be the escape rate from the well $q_1$ to the well $q_2$
and let $w_2$ be the escape rate in the opposite direction. 
It is well known that these escape rates are exponentially small,
$w_i\asymp\exp\bl\{-2(V^*-V_{q_i})/\eps\br\}$, where
$V^*$ is the value of the
maximum of $V$ on the interval $[q_1,q_2]$, if $\eps$ is small. 
Denote by $A_t$ 
the number of traders in the right well
$q_2$ at time 
$t\ge0$. Suppose that $A_0$ is much larger than $M$ implying that
the actual price at the initial moment is situated near $q_2$. We are
interested in describing the moment of the ``crash'', \ie\ when the
price moves from the right well $q_2$ to the left well~$q_1$. 

Then we can approximate the individual processes $p_i(t)$ by a
two-state
Markov chains with state space $\{q_1,q_2\}$ and transition rates
$w_1$ and $w_2$. In this approximation we can compute the
normalized expected number $a_t\equiv{\bf E} A_t/N$ of traders in state
$q_2$ at time $t$ as the solution to the ordinary differential equation
\[
\frac d{dt}a_t = -w_2a_t +w_1(1-a_t)\,.
\]
We get
\[
a_t=\frac{w_1}{w_2+w_1}+
\Bl(a_0-\frac{w_1}{w_2+w_1}\Br)\exp\bl\{-(w_2+w_1)t\br\}
\]
and the crash time $T_c$ can be defined as $t$ such that $a_t=M/N$,
\[
T_c=\frac1{w_2+w_1}\log
\frac{a_0(w_2+w_1)-w_1}{\frac MN(w_2+w_1)-w_1}
\]

If the energy barrier $\Delta V\equiv V^*-V_{q_2}$ is large enough,
the time for 
each single buyer to escape from the initial well is much larger than
the relaxation time for the system of $A_t$ particles in the right
well, and thus it
is natural to expect that the system will pass through the sequence of 
local equilibrium states corresponding to $A_t$ independent random
walkers. Using this observation, the evolution of the price can be
described in terms of the $B_t\equiv A_t-M$'s order statistic in
a system of 
$A_t$ random variables whose distribution is approximately
Gaussian with parameters $q_2$ and $\bl(V''(q_2)\br)^{-1}$.

To do this, let $F(\,\cdot\,)$ denote the distribution function of an
individual walker conditioned to stay near $q_2$ and define
$q_B=q(B_t)$ as the solution to the equation  
\[
\frac M{A_t}=1-F(q_B)
\approx 1-\Phi\bigl((q_B-q_2)\sqrt{ V''(q_2)}\bigr)\,,
\]
where $\Phi(\,\cdot\,)$ is the distribution function of the standard
Gaussian variable. Using the well-known asymptotics for the tail
distribution of $\Phi(\,\cdot\,)$, 
\[
\frac u{u^2+1}\frac{e^{-u^2/2}}{\sqrt{2\pi}}
\le\frac1{\sqrt{2\pi}}\int_u^\infty e^{-x^2/2}\,dx\equiv1-\Phi(u)
\le\frac1u\frac{e^{-u^2/2}}{\sqrt{2\pi}}\,,\qquad u\to\infty,
\]
we immediately get
\[
q_B\approx q_2-\Bigl(\frac{2\log(A_t/B_t)}{ V''(q_2)}\Bigr)^{1/2}
+\frac{\log(4\pi\log(A_t/B_t))}{\sqrt{8V''(q_2)\log(A_t/B_t)}},
\qquad B_t/M\to0\,.
\]
Consequently, 
in the limit of large $A_t$, $M$ such that $\rho=1-M/A_t$ is fixed we have
\[
q_B\approx q_2-\Bigl(\frac{2\log(\rho^{-1})}{ V''(q_2)}\Bigr)^{1/2}\,.
\]

\begin{figure}[th]
\centerline{\includegraphics[height=58mm,angle=270]{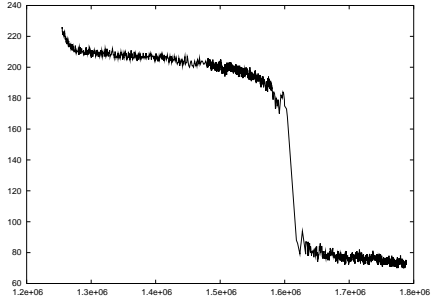}}
\caption{Crash in a double-well potential}
\end{figure}

In this regime of ``increasing ranks'', the fluctuations of the
$B_t$'s order statistic have Gaussian behaviour and their scaling can
be derived from~\cite[Theorem~2.5.2]{LLR}.
To do this, consider a small enough $y$ such that 
\[
A_t\,F(q_B+y)\bigl(1-F(q_B+y)\bigr)
\]
is large for large $B_t$ and such that
\[
\frac{M-A_t\bigl(1-F(q_B+y)\bigr)}{(B_t\,M/A_t)^{1/2}}\to\tau
\]
as $A_t$, $M=(1-\rho )A_t$, and $B_t=\rho A_t$ are getting large. In view of the
definition of $q_B$ the LHS expression above equals
\[
\sqrt{\frac{B_t\,A_t}M}\Bigl(\frac{F(q_B+y)}{F(q_B)}-1\Bigr)\,,
\]
where the last ratio can be approximated by
\[
\frac{q_2-q_B}{q_2-q_B-y}
\frac{\exp\bigl\{-(q_B+y-q_2)^2 V''(q_2)/2\bigr\}}
{\exp\bigl\{-(q_B-q_2)^2 V''(q_2)/2\bigr\}}
\approx1+(q_2-q_B) V''(q_2)\,y\,,
\]
assuming that $(q_2-q_B)^2 V''(q_2)$ is large enough. As a
result, in the limit of large $M$ and $B_t$, we
have 
\[
\tau\approx\frac{\sqrt{\rho M}(q_2-q_B) V''(q_2)}{1-\rho}\,y
\approx\frac{\sqrt{2\rho M V''(q_2)\log(\rho^{-1})}}{1-\rho}\,y\,.
\]
It remains to observe that Theorem~2.5.2 from \cite{LLR} implies then 
that the ask price $p^a(t)$ satisfies 
\[
\Pr\,\bl(p^a(t)
\le q_B+y\br)\to\Phi(\tau)\,.
\]
In other words, the price corresponding to a system of $A_t$ such
agents has mean $q_B$, shifted away from the well $q_2$ on a distance
of order $(\log(\rho^{-1})/V''(q_2))^{1/2}$ and the variance of the
price (ie, the 
volatility) diverges as $(MV''(q_2))^{-1}(\rho\log(\rho^{-1}))^{-1}$
in the limit of 
small $\rho=B_t/(M+B_t)$.

Finally, recalling the hydrodynamic description of $B_t$ and the
definition of $T_c$, we can also describe the time dependence of the
mean $\bfE\rho$,
\[
\bfE\rho=\bfE\rho_t\approx
\frac{1-\exp\{-(w_2+w_1)(T_c-t)\}}
{1+w_1\exp\{(w_2+w_1)t\}/\bl((w_2+w_1)a_0-w_1\br)}\,,
\qquad t\le T_c.
\]

\subsection{Interacting traders}

The simple model introduced in the previous section is rather
artificial and simplistic. In reality one would expect that the
behaviour of a trader is influenced by the information received from
the market as well as external influences. Moreover, the opinion held
by a trader with respect to the current price should somehow reflect
some his
 intrinsic psychological characteristics.
Finally,
the exchange of shares occurring when a transaction takes place should
have some visible effect on the time evolution. In the following we
suggest some minimal features that should be incorporated in an interacting 
model  to take into account some such
effects. We will see that these features correspond to types of interactions 
that are not commonly considered in the theory of interacting particle systems.

\begin{itemize}
 \item The derived process of the (ask and bid) price is the most easily 
   accessible piece of information about the trading state of the 
   market for any trader. It is natural that the updating rules should take
the current value of this process into account. 
 The simplest and natural modification
is to introduce a
bias towards the actual price
 $\bl(p^a(t),p^b(t)\br)$ into the distribution of opinion change.
 \item Traders whose opinion is far from the current price are likely not
to pay much attention to what is happening  on  the market. It is reasonable 
to assume that they update their opinion less frequently. 
This feature can be included by reducing the overall
transition rates as a function of $p_i(t)-p(t)$. 
 \item Finally, it is natural to assume that the traders performing a
 transaction, that is exchange of a share, will update their opinions
 according to some special rules reflecting the fact that someone
 buying or selling a share at a given price believes that she has
 struck a favourable deal, \ie\ they attribute a higher value to the
 share then what they paid, respectively a lower one then what they
 got. 
\end{itemize}

In the following we describe some concrete framework 
in which these features are implemented. We describe the 
construction of the process algorithmically. 

{\it Change of opinion:} 
 At any time step we first select at random a
trader. We will allow this probability to depend on $p_i$, and
  we 
choose trader $i$ with a probability proportional to $f(p_i(t)-p(t))$,
where we define the ``current price'' via
$p(t)=\bl(p^a(t)+p^b(t)\br)/2$ and the function 
$f(x)\geq 0$ has its maximum at zero. The function $f$ is responsible
for the slow-down phenomenon away from $p(t)$. Once a trader has been
selected, she changes her opinion from $p_i$ to $p'_i$ with
probability proportional to $q(p_i,p'_i)$ which in turn may depend on
the entire state of the system. A possible choice for these functions is 
\[
f(x)=1/(1+|x|)^\alpha, \qquad \alpha>0\,,
\]
and $q(x,y;\bfp)$ being, for any fixed $\bfp$, a kernel of a random
walk. In typical cases, $q(x,y;\bfp)$ depends on $\bfp$ through
$p^a$ and $p^b$ only. Once $p'_i$ is chosen, we check whether
$p'_i<p^a$, if $n_i=0$, resp. whether  $p'_i>p^b$, if $n_i=+1$. If
this is the case, we set $p_i(t+1)=p_i'$, and $p_j(t+1)=p_j(t)$ for
all $j\neq i$, and continue to the next time step. Otherwise, we
perform  

{\it Transaction}: Assume first that $n_i(t)=0$ and
$p'_i\geq p^a(t)$. This means that the buyer $i$ has decided to buy at
the current asked price. Since by definition there is at least one
seller who asks only the price $p^a(t)$, we select from all these one
at random with equal probabilities. Call this trader $j$. Then we set 
\[
p_i(t+1)=p^a(t)+g,\qquad
p_j(t+1)=p^a(t)-g
\]
where $g>0$ is a fixed or possibly random number. Similarly, if
$n_i(t)=1$ and $p'_i\leq p^b(t)$, the seller $i$ sells to one of the
buyers that offer the price $p^0(t)$, and we set
\[
p_i(t+1)=p^o(t)-g,\qquad
p_j(t+1)=p^o(t)+g
\]
The final state in all cases represents a new stable trading  state and the 
process continues. Note that $g$ should be at least as large as to
cover the transaction cost.  

These additional features make the  mathematical analysis of the
model much more difficult,  but they introduce some interesting
effects that are somewhat similar to phenomena observed in real 
markets. Let us briefly comment on these.

In the ideal gas approximation in the
absence of any confining potential
all opinions would in the long run spread over all real numbers
 and 
the individual
opinions would  get arbitrarily far from each other. 
This is avoided by the mechanism of the {\em attraction to the current price}.

\begin{figure}[h]
\centerline{\includegraphics[height=58mm,angle=270]{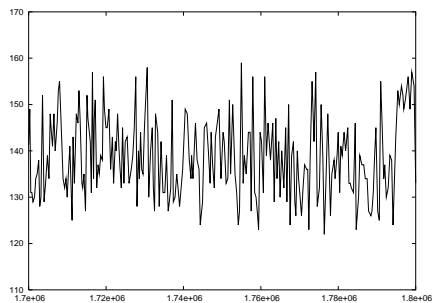}}
\caption{Trading in between resistances}\label{resistances}
\end{figure}

{\em Slowing down} the jump rate of the particles far away from the current
price naturally introduces long time  memory effects that lead to
special features in the distribution of the price process. One of them
is possible existence of resistances: if there is a large population
of traders in a vicinity of some price $p$ far from (say above) the
current price $p(t)$, then this population tends to persist for a long
time, unless the current price approaches this value. If that happens,
\eg\ due to the presence of an upward drift, one observes a slow-down
of the upward movement of the price
when it 
approaches this value. The
market has a resistance against increase of the price through this
value, reflected in multiple returns to essentially the same extremal
values for a long period, see Fig~\ref{resistances}.

\begin{figure}[h]
\centerline{\includegraphics[height=58mm,angle=270]{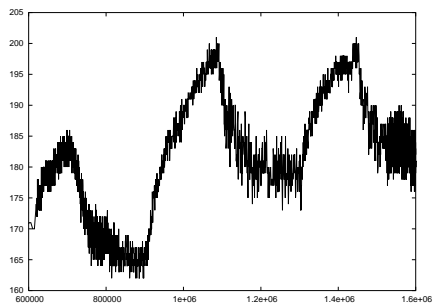}}
\caption{A sequence of ``bubbles''}\label{bubbles}
\end{figure}

A further effect of the slowing down 
is the tendency of the creation of ``bubbles'' in the
presence of strong drifts. In this case one observes a fast motion of
the price accompanied by a depletion of the population below this
price. Effectively, a few (buying) traders move with the drift, while
most are left behind. Such a situation can lead to a crash, if at some
moment the drift is removed (due to external effects). Such a scenario
was played out in a simulation that is shown in Figure~\ref{bubbles}.

Note that after the crash there was a strong 
increase of volatility. 

The effect of pushing the opinions from the current price after the 
{\em transaction } depletes the vicinity of the price and therefore 
increases the volatility. This effect goes in the opposite direction 
as the attraction to the price and the interplay of both effects can 
lead to a non-trivial quasi-equilibrium state. 
The effect of these 
mechanisms on the price fluctuation will be studied in
forthcoming 
paper \cite{BCH}.

\bigskip
\begin{figure}[h]
\centerline{\includegraphics[height=58mm,angle=270]{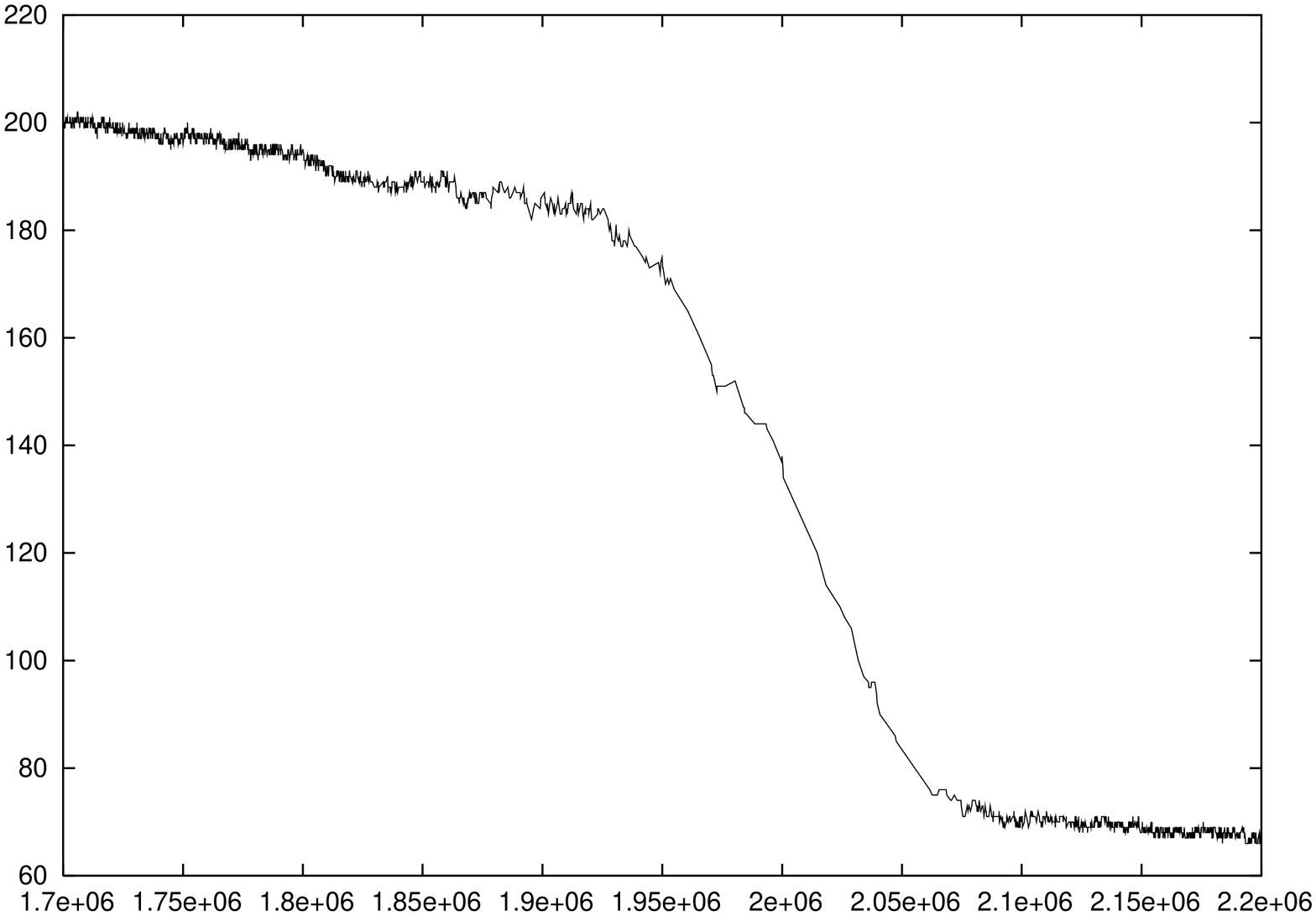}\qquad%
\includegraphics[height=58mm,angle=270]{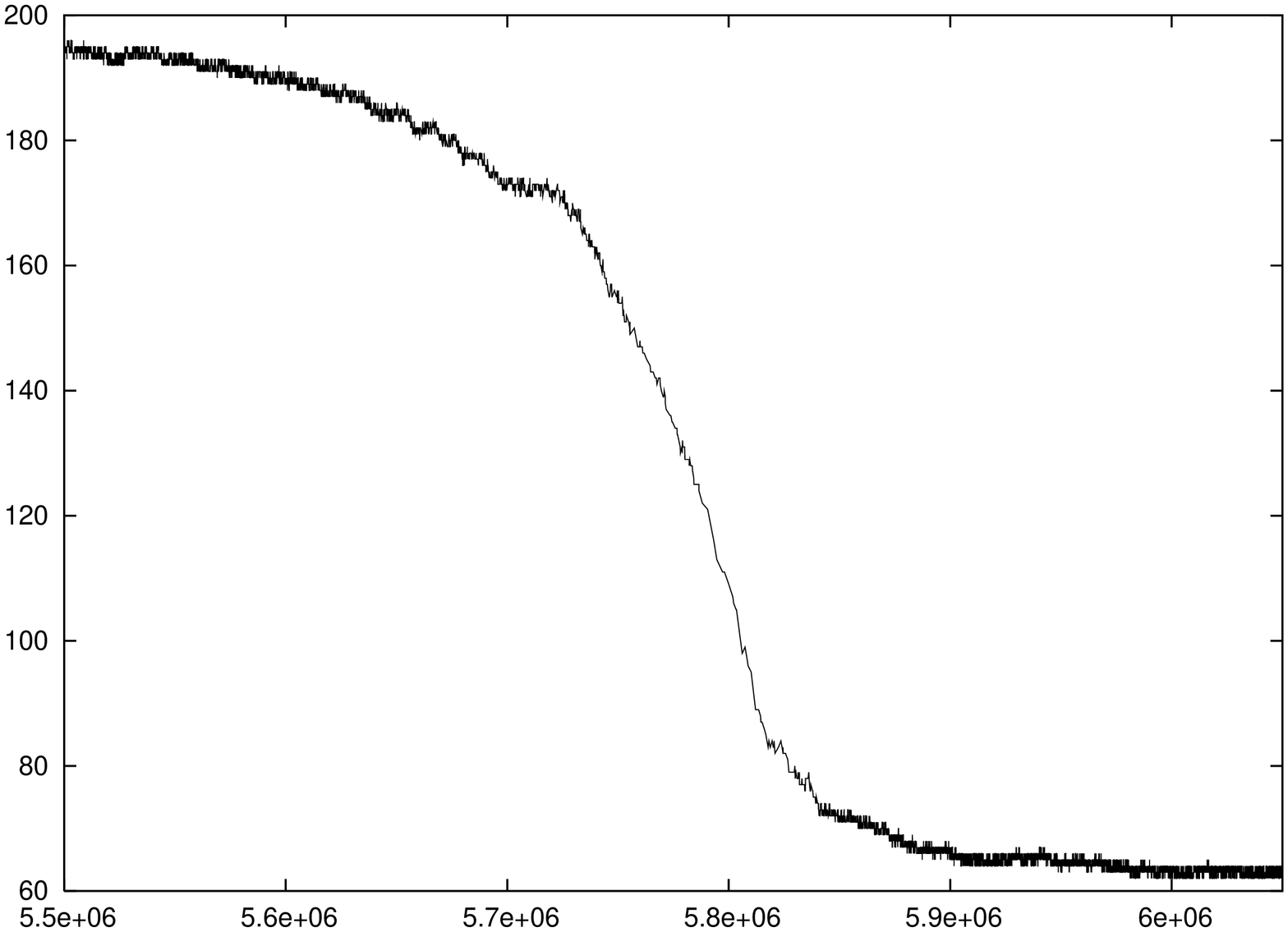}}
\vskip1cm
\centerline{\includegraphics[height=58mm,angle=270]{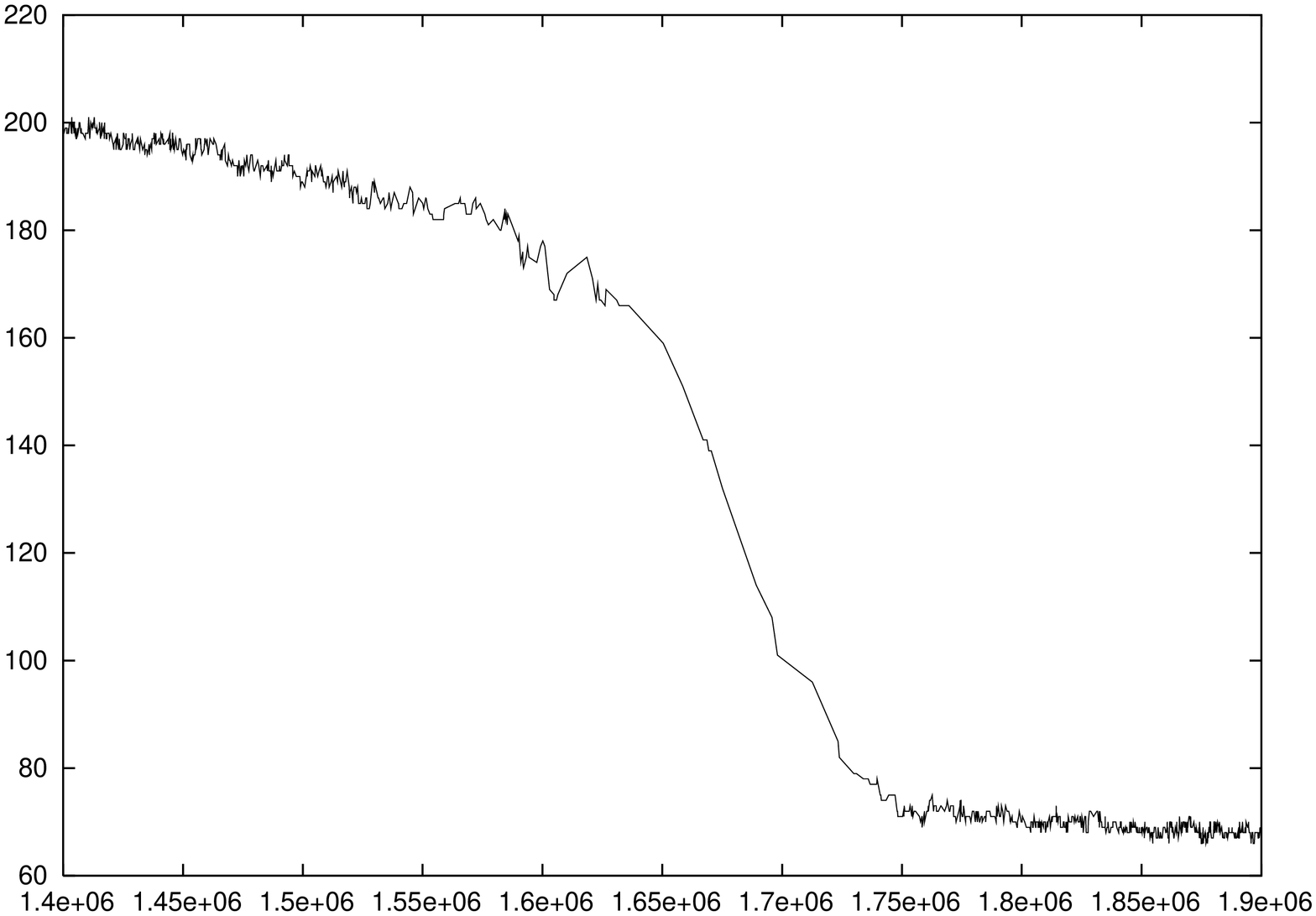}\qquad%
\includegraphics[height=58mm,angle=270]{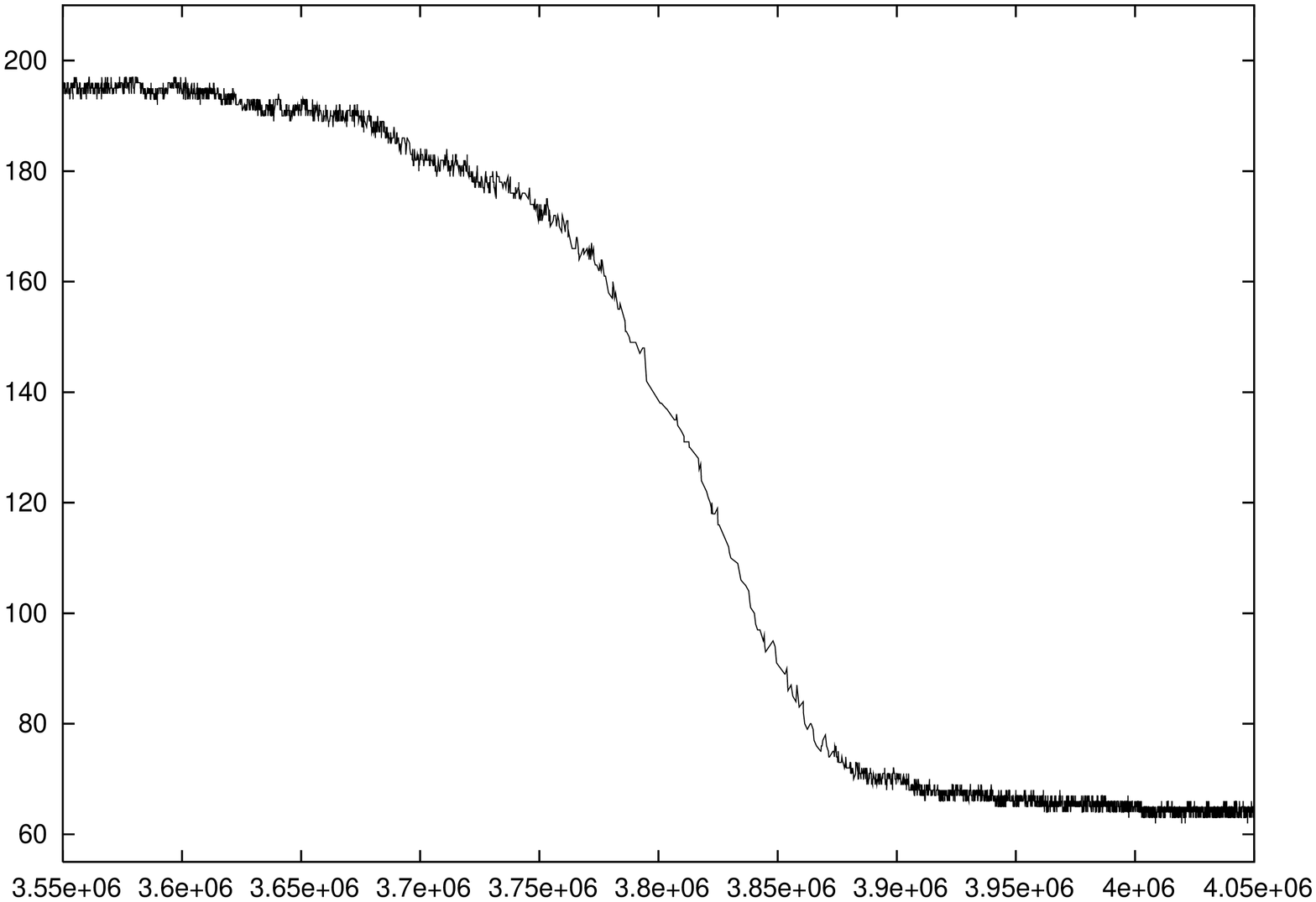}}
\figtext{ 
\writefig       18    4.5     {1)} 
\writefig       64    4.5     {2)} 
\writefig       18    -.5     {3)} 
\writefig       64    -.5    {4)} 
}
\vskip12mm
\caption{Crash scenario without slowdown}\label{NOslowdown}
\end{figure}
\begin{figure}[h]
\centerline{
\includegraphics[height=58mm,angle=270]{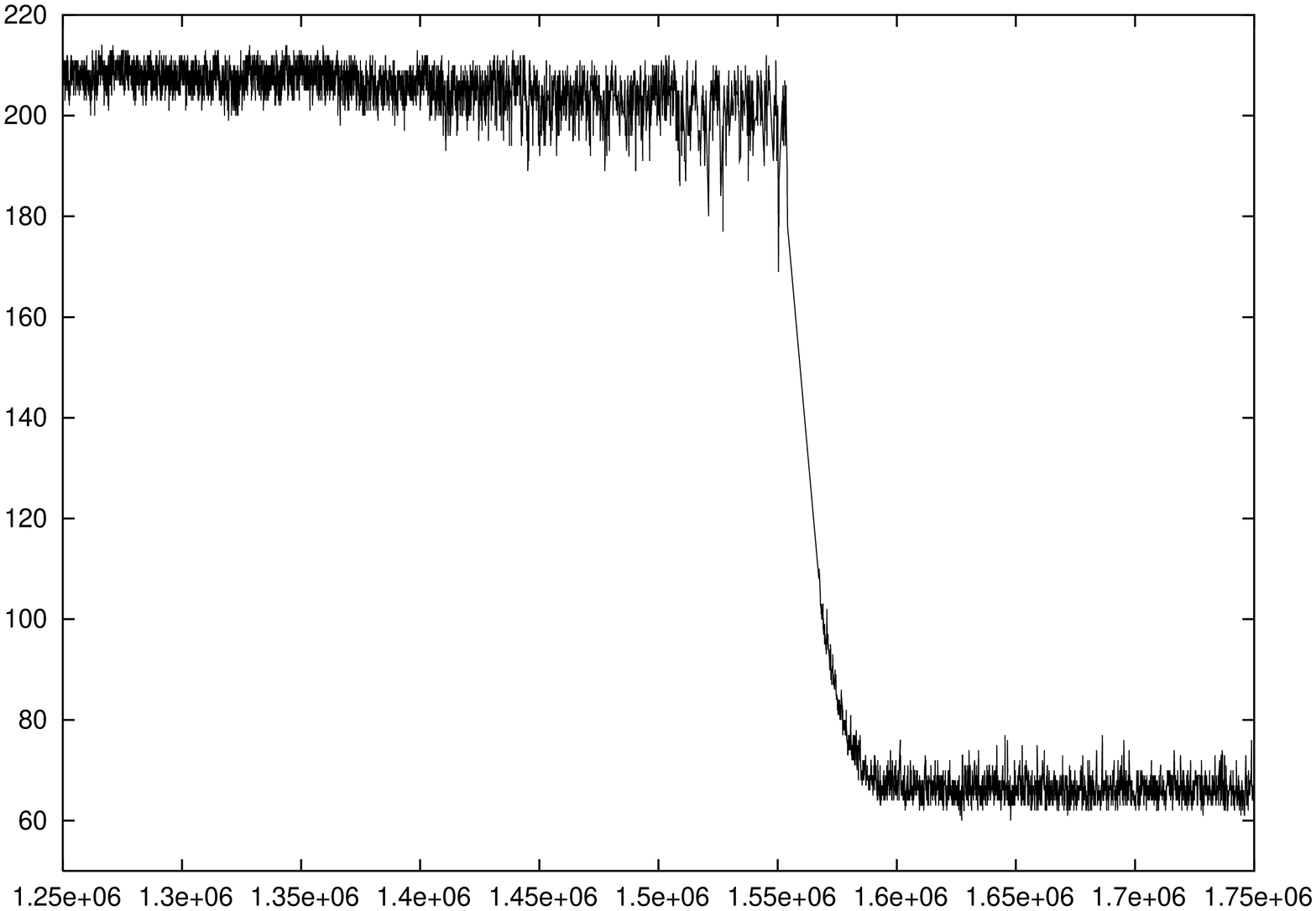}\qquad%
\includegraphics[height=58mm,angle=270]{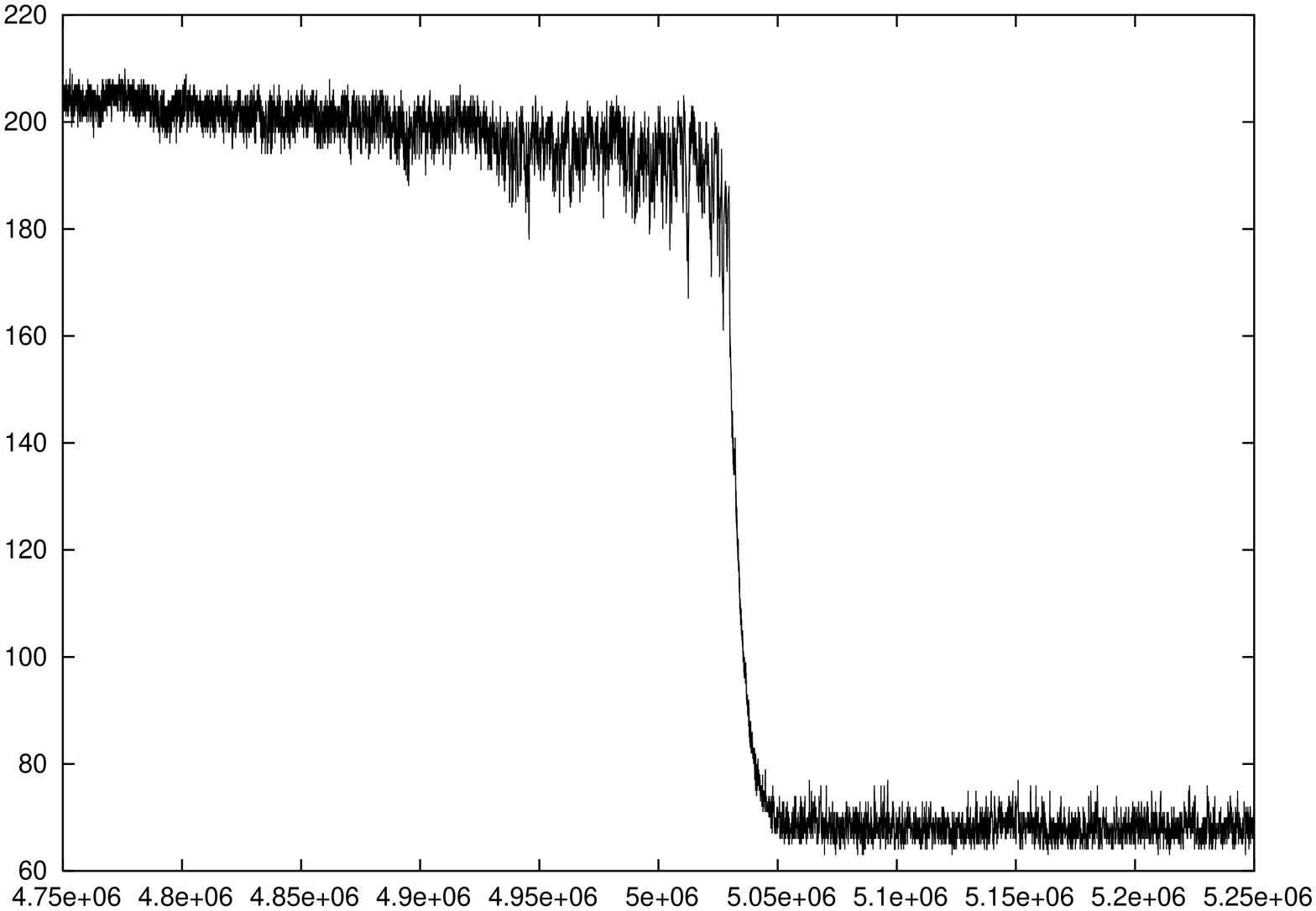}}
\vskip1cm
\centerline{
\includegraphics[height=58mm,angle=270]{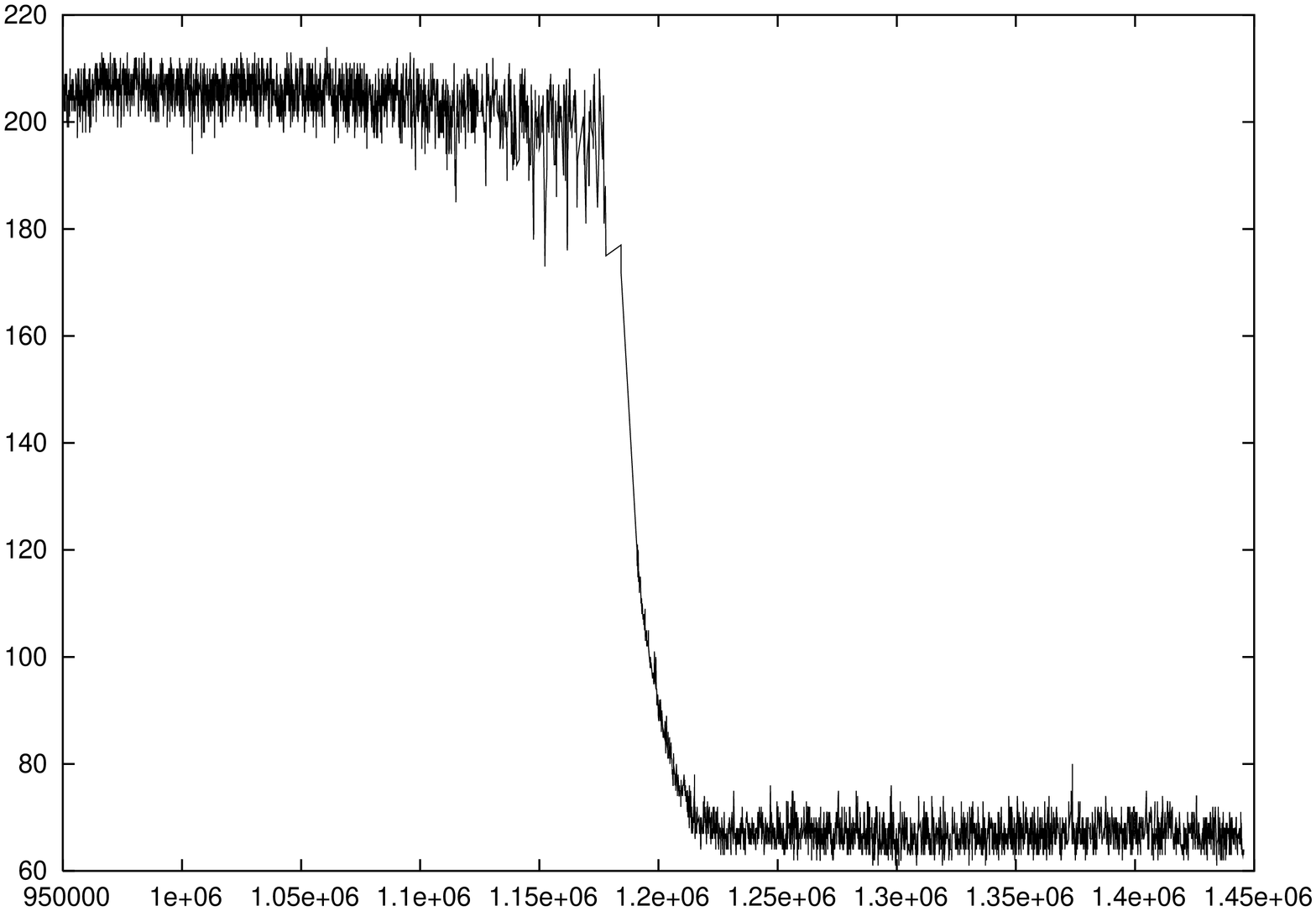}\qquad%
\includegraphics[height=58mm,angle=270]{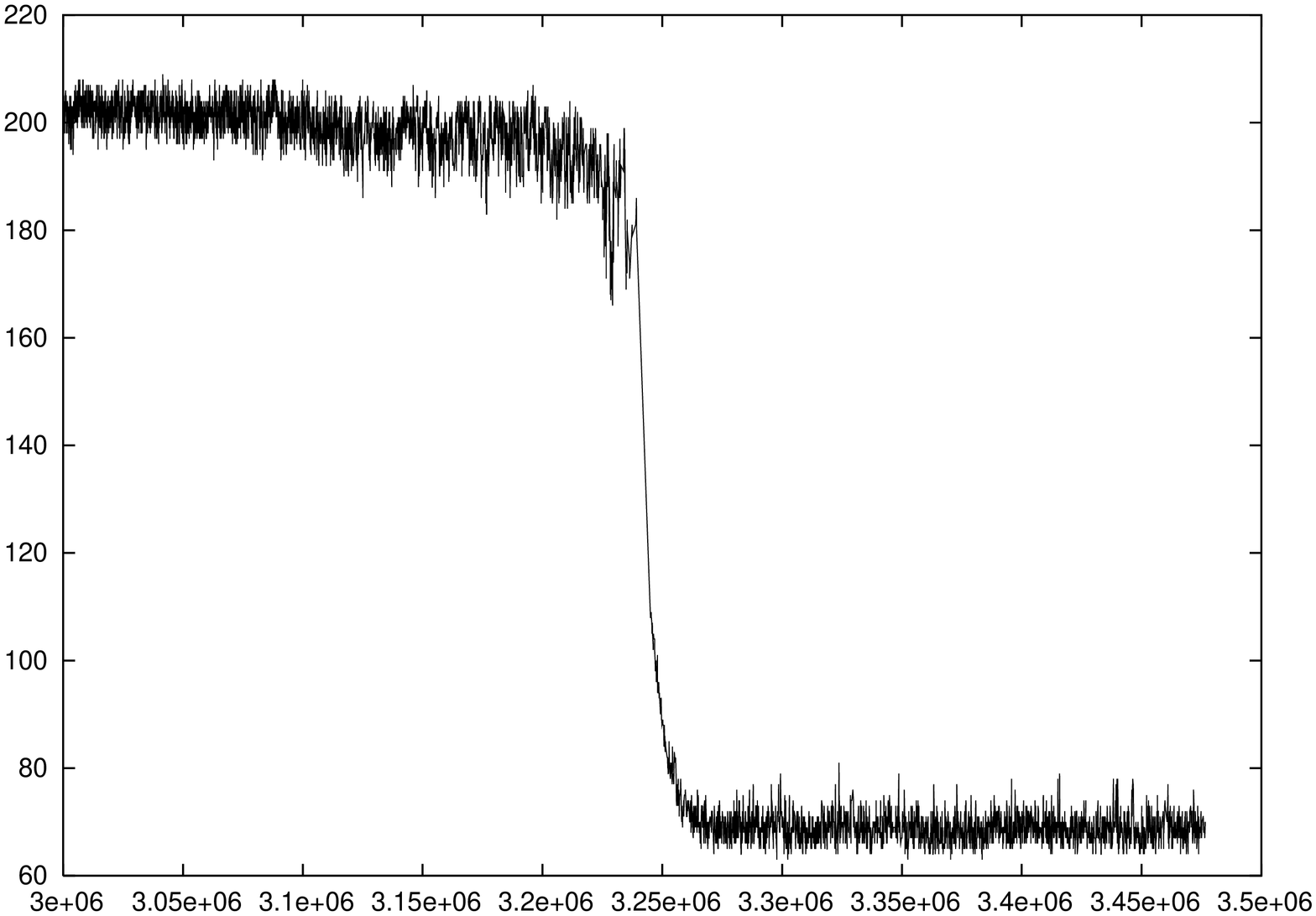}
}
\figtext{ 
\writefig       18    4.5     {5)} 
\writefig       64    4.5     {6)} 
\writefig       18    -.5     {7)} 
\writefig       64    -.5    {8)} 
}
\vskip12mm
\caption{Crash scenario with slowdown}\label{slowdown}
\end{figure}

To illustrate the influence of different effects discussed above, we
present in Fig.~\ref{NOslowdown} and~\ref{slowdown}
simulation results of the crash-type scenario from
Sect.~3.1 where three different parameters of the model were changed
independently. All simulations are based on a discretized version of the 
diffusion model from Sect.~3.1 with the same potential function $V$ 
having two local minima (one metastable and one stable). All simulations start 
with the same initial condition where all traders are located near the 
metastable minimum.

Figure~\ref{NOslowdown} shows simulations without the feature of
slowing down rates as a function of the distance to the current price: 
1) corresponds to the free gas approximation; 
2) shows the same
scenario with an additional drift towards the current price; 3) and 4)
are like 1) and 2) with a trading effect corresponding to 
$g$ being uniformly distributed on the interval~$[3,10]$. 

Figure~\ref{slowdown} shows the same sequence of scenarios when the
overall rate of updating of trader $i$ behaves like
$\bl(p_i-p(t)\br)^{-1.5}$. Notice the increased volatility compared to the
previous picture. The volatility increase before the crash is
particularly marked.

\section{Possible extensions\label{four}}

The basic model we describe above allows for numerous extensions to capture
further important features. 

\subsection{Traders of different type.}

We have assumed all traders to behave according to the same stochastic rules.
It is not difficult to modify this.   First, rules can be different between 
buyers and sellers, generalizing the bias towards the price to some more 
complicated function. Moreover, we could introduce different species of 
traders that follow different rules (\eg\ optimists \vs\ pessimists),
and study  
the ensuing effects. Even more challenging, we could try to introduce 
``intelligent agents''  that try to perform arbitrage on the price by 
estimating future price based on observation of the past price history.
This goes beyond the original ideas of the model, but  could be interesting
when testing some basic principles of financial mathematics in a specific 
controllable model context.

\subsection{Coupling to external influences.}

To capture  the evolution of prices over longer time-scales, 
it will be indispensable to couple our model to external influences. 
These should reflect fundamental data on the particular stock considered
(such as dividend return, earnings, cash-flow), as well as global 
macro-economic data (interest rates, growth rates, etc.). Such effects are 
easily incorporated by making the transition rates $q(p_i,p_i')$
time dependent. \Eg, given the earnings at time $t$, one may compute
a fictitious ``fundamental value'' (based \eg\ 
on historic 
price-earnings ratios), and assume that there should be a certain tendency
for market participants to adjust their subjective price 
towards this value. Changes in earnings (expectations) then induce a change in 
the transition probabilities. Similarly, other external effects
exert their influence most naturally through the transition probabilities
of our process. The key question of interest that our model is able to 
answer is how such external effects are reflected in the evolution of the
price of our commodity. Addressing this question via analytical and/or 
numerical methods may in  many respects  be the most interesting 
and promising perspective that our models provide. 

\section{Conclusions\label{five}}
We have presented a class of Markov models that allow a realistic
modelling of the price evolution of a commodity under trading. The
basic model is a particle system like model for the evolution of a
large number of traders whose state space is given by the collection
of all opinions of all traders on the current value of the traded
commodity. The price process is inferred from this state according to
rules analogous to those used in real markets. In the simplest case
of independent traders, explicit computations are possible, and we
have analysed a crash scenario in a bistable market in this context.  

We discussed several basic mechanisms 
that we think 
should be taken into account when modelling 
financial markets. These include attraction to the price, rate
dependence from the  
distance to the price, and repulsion from the price of traders having 
performed an interaction. These effects lead to interesting properties of the
price process which are observed in similar form in reality. 

We hope to have motivated that the interacting particle systems have a place in
 the modelling of financial and economic systems. For this to be fruitful, this
requires to chose interactions that take the special features of
these systems into account. Moreover, the questions that should be addressed 
are quite different from what is usually done in the theory of interacting
particle systems. In particular the analysis of the price process leads to
rather interesting problems regarding order statistics in the particle model.
As we will discuss in a
forthcoming article \cite{BCH}, these problems
 are closely related to the study of interfaces and phase boundaries
 in particle systems.  

\section*{Acknowledgments}
A.B. and J.C. are supported in part by the DFG-Research Center ``Mathematics 
for key technologies''. O.H. thanks the Weierstra\sz\ Institute, Berlin, for
its kind hospitality and the DFG-Research Center ``Mathematics 
for key technologies'' for financial support. A.B. thanks the Isaac
Newton Institute, Cambridge (UK), for its kind hospitality and financial
support within the programme ``Interaction and growth in complex
stochastic systems''.


\end{document}